\documentclass[reprint,aps,pra,twoside,superscriptaddress,nofootinbib,longbibliography]{revtex4-2}
\pdfoutput=0

\usepackage{graphicx}% Include figure files
\usepackage{dcolumn}% Align table columns on decimal point
\usepackage{bm}% bold math
\usepackage{multirow}
\usepackage{amsmath,amsfonts,mathtools,tensor}
\usepackage{amssymb,latexsym}
\usepackage{color}
\usepackage{mathrsfs}
\usepackage{comment}
\usepackage{siunitx}
\usepackage[normalem]{ulem}

\newtheorem{prop}{Proposition}

\newcommand{\beprop}{\begin{prop}}
\newcommand{\enprop}{\end{prop}}
\newcommand{\bprf}{\begin{proof}} 
\newcommand{\eprf}{\end{proof}\qed}

\definecolor{hervecolor}{rgb}{0.8,0,0.7}

\newcommand{\ket}[1]{|\kern.3ex#1\kern.3ex\rangle}
\newcommand{\bra}[1]{\langle\kern.3ex #1 \kern.3ex|}
\newcommand{\scalar}[2]{\langle\kern.3ex #1 \kern.3ex|\kern.3ex#2\kern.3ex\rangle}

\newcommand{\ii}{\mathsf{i}}

\newcommand{\ud}{\mathrm{d}}

\newcommand{\uE}{\mathrm{E}}
\newcommand{\uD}{\mathrm{D}}

\usepackage[colorlinks]{hyperref}
\hypersetup{
     breaklinks=true,
    pdfstartview={FitH},    % fits the width of the page to the window
    colorlinks=true,       % false: boxed links; true: colored links
    linkcolor=blue,          % color of internal links
    citecolor=red,        % color of links to bibliography
    filecolor=magenta,      % color of file links
    urlcolor=blue,           % color of external links
    anchorcolor=green,      % Color for anchor text
    linktocpage=true
}
\usepackage{cleveref}

\begin{document}

%\preprint{APS/123-QED}

\title{Torsion-Driven Nonlinearity in Spinless Quantum Mechanics}% Force line breaks with \\

\author{Tomoi Koide}
\email{tomoikoide@gmail.com,koide@if.ufrj.br}
\affiliation{Instituto de F\'{\i}sica, Universidade Federal do Rio de Janeiro, 
21941-972, Rio de Janeiro, RJ, Brazil}
\affiliation{Frankfurt Institute for Advanced Studies (FIAS), Frankfurt am Main, Germany}  
\author{Armin van de Venn}
\email{venn@fias.uni-frankfurt.de}
\affiliation{Frankfurt Institute for Advanced Studies (FIAS), Frankfurt am Main, Germany}
\affiliation{Physics Department, Goethe University, Max-von-Laue-Str.~1, 60438 Frankfurt am Main, Germany}


\begin{abstract}
We investigate the previously unexplored quantum dynamics of
non-relativistic, spinless particles propagating in curved spaces with
torsion.
Our findings demonstrate that while torsion has been predominantly
associated with spin, it can also influence the quantum behavior of
spinless particles by inducing a logarithmic nonlinearity in the
Schr\"{o}dinger equation through quantum fluctuations, even in flat space.
To facilitate quantization in curved spaces, we introduce an extended stochastic variational method. Unlike canonical quantization, this
approach is naturally suited to general coordinate systems, with quantum
fluctuations arising from a noise term in the stochastic process that is
directly influenced by torsion. By requiring consistency with quantum
dynamics, we derive an upper bound on the magnitude of torsion.
Our results reveal a previously unrecognized mechanism by which torsion,
as predicted in certain extensions of general relativity, can influence
quantum systems, with potential implications for early-universe physics
and dark matter or energy models.
\end{abstract}

\maketitle

\section{introduction}

General Relativity (GR) describes gravity as a geometric property of spacetime and has successfully explained various astrophysical and cosmological phenomena, 
such as gravitational lensing and the expansion of the universe. 
However, standard GR-based models face difficulties in incorporating additional gravitational effects, suggesting the need for modifications to our understanding of gravity.

One promising extension is the metric-affine gravity theory, which treats the connection as an independent field, introducing additional degrees of freedom: torsion and nonmetricity. 
Torsion accounts for the noncommutativity of parallel transport in spacetime, 
while nonmetricity quantifies the extent to which the connection fails to preserve metric compatibility. 
This type of theory addresses the limitations of GR in explaining phenomena like dark matter and dark energy \cite{Belyaev16,delaCruzDombriz21,Kirsch23,vandeVenn22}, offering a broader framework to explore gravitational effects at both classical and quantum scales \cite{BeltranJimenez19,Percacci20}.
Torsion has predominantly been studied in the context of spin interactions in particle physics, where it naturally arises in the Dirac equation \cite{Hehl2, Hehl1, BAUERLE83, Sabbata94, SHAPIRO02, Poplawski10,vandeVenn24}, yet its role in quantum fluctuations within spinless systems remains 
{\it terra incognita}.

In this paper, we demonstrate that torsion influences quantum mechanics even in the absence of spin, suggesting a previously overlooked connection between spatial geometry and quantum fluctuations. 
To establish this, we extend the stochastic variational method (SVM) 
\cite{NELSON, YASUE81, ZAMBRINI, Matos20, Kuipers} to spaces with torsion. This framework is particularly well-suited for our investigation, as it provides a physically intuitive picture of quantization. 
In SVM, quantum effects are not postulated abstractly but are seen as emerging directly from the stochastic, Brownian motion-like trajectories of particles. 
For instance, the Schr\"{o}dinger equation emerges naturally when SVM is applied to the classical Newtonian action, with position-momentum uncertainty stemming from the inherent non-differentiability of particle trajectories \cite{KOIDE20181472,GAZEAU20,Koide_2022_JSTAT}.  
Unlike canonical quantization, which is limited to variables with unbounded spectra, such as Cartesian coordinates, 
SVM accommodates generalized coordinates like angular variables \cite{GAZEAU20,KOIDE20192713,Koide_2020}. 
This unique feature allows us to study the central question of this paper: how a geometric property like torsion, which does not appear in the classical Lagrangian, can nevertheless alter quantum dynamics by directly influencing the nature of the underlying stochastic fluctuations.

This paper is organized as follows. 
In Sec. \ref{sec:torsion}, we develop the mathematical framework to describe a curved space with torsion. 
In Sec. \ref{sec:svm}, we extend SVM to be applicable in this geometric setting. 
In Sec. \ref{sec:nonlinearSCH}, we apply this extended SVM to a non-relativistic particle to derive its quantum dynamics. 
Finally, in Sec. \ref{sec:conclusion}, we present concluding remarks.

\section{Curved spaces with torsion} \label{sec:torsion}

We adopt the vielbein formulation of differential geometry, which offers a convenient framework for working in curved spaces \cite{book:vasak}. 
Consider a $D$-dimensional curved space equipped with a metric tensor $g_{ij}$ in a coordinate basis, where $i, j = 1, 2, \ldots, D$. 
In contrast, in a local orthonormal frame, the metric is expressed as $\eta_{\hat{a}\hat{b}}$, with vielbein indices $\hat{a}, \hat{b} = 1, 2, \ldots, D$. 
Throughout this paper, we use a hat symbol (i.e., $\,\hat{\bullet}\,$) to indicate components in the local orthonormal frame.

The coordinate and orthonormal representations of the metric are related via the vielbein $\tensor{e}{^{\hat{a}}_i}$ according to
\begin{equation}
g_{ij} = \eta_{\hat{a}\hat{b}} \tensor{e}{^{\hat{a}}_i} \tensor{e}{^{\hat{b}}_j} \, . \label{eqn:g-e}
\end{equation}
Together with its dual vielbein 
$\tensor{e}{_{\hat{a}}^{i}} \coloneqq g^{ij} \eta_{\hat{a}\hat{b}} \tensor{e}{^{\hat{b}}_j}$, the vielbein satisfies the orthonormality relations 
$\delta^i_j
= \tensor{e}{_{\hat{a}}^{i}} \tensor{e}{^{\hat{a}}_j}$
and 
$\delta^{\hat{a}}_{\hat{b}} 
= \tensor{e}{^{\hat{a}}_{i}} \tensor{e}{_{\hat{b}}^{i}}$.
Parallel transports expressed in coordinate and vielbein indices are, 
respectively, characterized by the affine connection $\Gamma$ and spin (Lorentz) connection $\omega$.
The latter is defined as
\begin{align}
{{\omega}^{\hat{a}}}_{\hat{b}i} 
\coloneqq 
%e^{\hat{a}} \nabla_i e_{\hat{b}} 
%=
 \tensor{e}{^{\hat{a}}_j} \partial_i \tensor{e}{_{\hat{b}}^{j}} 
+ \tensor{e}{_{\hat{b}}^j}  {\Gamma^k}_{j i}  \tensor{e}{^{\hat{a}}_{k}} 
\, .
\end{align}
Henceforth, we assume metric compatibility $\nabla_k g_{i j} = 0$,
leading to the antisymmetry of the spin connection 
in its first two indices
$\omega_{\hat{a} \hat{b} i}
= -\omega_{\hat{b} \hat{a} i}$.

In the presence of torsion, the affine connection can be decomposed as
\begin{equation}
    \Gamma\indices{^{k}_{ij}} = \mathring{\Gamma}\indices{^{k}_{ij}} + K\indices{^{k}_{ij}} \, ,
    \label{eqn:contorsion}
\end{equation}
where $\mathring{\Gamma}\indices{^{k}_{ij}}$ represents the Levi-Civita connection, 
and $K\indices{^{k}_{ij}}$ denotes the contortion tensor,
\begin{equation}
    K\indices{^{k}_{ij}} \coloneqq 
    g\indices{^{kl}} \left( S\indices{_{lij}} - S\indices{_{ilj}} - S\indices{_{jli}} \right) \, ,
\end{equation}
which is defined in terms of the torsion tensor,
\begin{equation}
    S\indices{^{k}_{ij}} \coloneqq \Gamma\indices{^{k}_{[ij]}} = \frac{1}{2} \left( \Gamma\indices{^{k}_{ij}} - \Gamma\indices{^{k}_{ji}} \right) \, .
\end{equation}
We will explore how this additional degree of freedom, represented by torsion, influences quantum dynamics.

\section{Extended stochastic variational method}
\label{sec:svm}

In the SVM quantization, quantum particles follow a Brownian motion where noise embodies quantum fluctuations. 
For a $D$-dimensional curved space with torsion, we extend this formulation to capture the quantum dynamics.
The system is defined by a time-independent Riemannian metric $g_{ij}$, where the particle position 
$\widetilde{q}^i (t)$ obeys a forward stochastic differential equation (SDE) with $\ud t > 0$ \cite{ito,ZAMBRINI,KOIDE20192713}:
\begin{equation} 
\ud \widetilde{q}^i (t) = 
u^i_+ (\widetilde{q}(t),t) \ud t + \sqrt{2\nu} \tensor{e}{_{\hat{a}}^i} \circ \ud\widetilde{B}^{\hat{a}}_+ (t) \, . \label{eqn:fSDE}
\end{equation} 
Here, $\, \widetilde{\bullet}\, $ denotes stochastic quantities, $\ud \widetilde{A}(t) := \widetilde{A}(t+\ud t) -\widetilde{A}(t)$ for an arbitrary stochastic quantity $\widetilde{A}(t)$ and $\nu$ characterizes the noise intensity. 
This is a standard stochastic differential equation (Langevin equation) describing Brownian motion in a curved space.
The first term on the right-hand side represents drift, and the second term represents noise.
Due to the noise term, the particle's trajectory is always zigzag, and the non-differentiability of this trajectory is evidence of quantum effects, which is the idea behind quantization using SVM.
The smooth vector field $u^i_+(q,t)$ determined by stochastic fluctuations is not the particle's velocity itself.
See also Eqs.\ (\ref{eqn:q_forwardderi}) and (\ref{eqn:q_backwardderi}).
This equation makes clear that the vielbein included in the noise term affects (quantum) fluctuations.
The inclination of the Wiener process $\ud\widetilde{B}^{\hat{a}}_+ (t)$ 
satisfies the following correlations, 
\begin{align}
\uE\left[\ud \widetilde{B}^{\hat{a}}_+ (t) \right] &= 0 \, , \\
\uE\left[\ud \widetilde{B}^{\hat{a}}_+ (t) \, \ud \widetilde{B}^{\hat{b}}_+ (t^\prime) \right] &= \ud t \delta^{\hat{a} \hat{b}} \delta_{t t^\prime} \, ,
\end{align}
where $\uE[\, \bullet \, ]$ denotes the ensemble average.
These are defined in the local Cartesian coordinates and 
projected onto the curved space 
using the vielbein $\tensor{e}{_{\hat{a}}^i} = \tensor{e}{_{\hat{a}}^i} (\widetilde{q}(t))$. 
The final ingredient of Eq.\ \eqref{eqn:fSDE} is the so-called
Stratonovich product, denoted by symbol $``\circ"$, 
which is defined for arbitrary smooth functions 
$F(q,t)$ and $G(q,t)$ as follows:
\begin{align}
& F(\widetilde{q}(t),t) \circ \ud G(\widetilde{q}(t),t) \nonumber \\
&\coloneqq \frac{F(\widetilde{q}(t+\ud t),t+\ud t)+ F(\widetilde{q}(t),t)}{2} \ud G(\widetilde{q}(t),t) \, .
\end{align}
To express Eq.\ (\ref{eqn:fSDE}) 
without the Stratonovich product, 
we need the difference of the vielbein. 
Following It\^{o}'s definition of stochastic parallel transport \cite{ito}, we adopt 
\begin{equation} 
\tensor{\ud e}{_{\hat{a}}^i} 
:= -  {\Gamma^{i}}_{j k}  \tensor{e}{_{\hat{a}}^j}  \circ  \ud \widetilde{q}^k(t) \, . \label{eqn:de}
\end{equation}

To define the stochastic variation, we need to introduce a derivative on its non-differentiable, zigzag trajectory.
As shown later in Eqs.\ (\ref{eqn:mfd}) and (\ref{eqn:mbd}), it requires the introduction of a time-reversed, backward SDE. To derive this backward SDE consistently, however, we must first establish the Fokker-Planck (FP) equation that governs the particle's probability distribution.
The invariant conserved particle probability distribution 
is defined by 
\begin{equation}
\rho ({q},t)= \frac{1}{\sqrt{g(q)}} \int  \ud^D R \sqrt{g(R)} \, \rho_0 (R) \uE \left[ \delta^{(D)} ({q}-\widetilde{q}(t)) \right] \, ,
\end{equation}
where 
$\delta^{(D)} (q - \widetilde{q}(t)) = \prod_{i=1}^D \delta (q^{i} - \widetilde{q}^{i}(t))$, $\rho_0(q) = \rho(q,0)$, 
$R^{i} = \widetilde{q}^{i} (0) = q^{i} (0)$ and $g = {\rm det} (g_{ij})$. 
Applying Itô's lemma \cite{gardiner}, the differential of the delta function is given by 
\begin{align}
& \ud \delta^{(D)} (q-\widetilde{q}(t)) \nonumber \\
&= 
\left[ 
\ud \widetilde{q}^i (t) \partial_{\widetilde{q}^i} + \frac{\ud \widetilde{q}^i (t) \ud \widetilde{q}^j (t)}{2} \partial_{\widetilde{q}^i} \partial_{\widetilde{q}^j} 
\right] \delta^{(D)} ({q}-\widetilde{q}(t)) + o ((\ud t)^{3/2})\nonumber \\
&= 
\left[ 
\ud \widetilde{q}^i (t) \partial_{\widetilde{q}^i} + \nu (\ud t) g^{i j} \partial_{\widetilde{q}^i} \partial_{\widetilde{q}^j} 
\right] \delta^{(D)} ({q}-\widehat{q}(t)) + o ((\ud t)^{3/2}) \, ,
\end{align}
Here, $\ud \widetilde{q} (t)$ is expressed in its It\^{o} form, which is derived from the Stratonovich form in Eq.\ (\ref{eqn:fSDE}): 
\begin{equation} d \widetilde{q}^i (t) = \left( u^i_+ (\widetilde{q}(t),t) - \nu 
\Gamma\indices{^{i}_{k j}} g^{k j} \right) dt + \sqrt{2\nu} \tensor{e}{_{\hat{a}}^i}  d\widetilde{B}^{\hat{a}}_+ (t) \, . 
\end{equation} 
To obtain the FP equation, we substitute this expansion into the definition of the probability density and take the ensemble average, utilizing the identity: 
\begin{align}
 \partial_k \, g^{ik} \sqrt{g}
= 2 g^{il} \sqrt{g} {S^k}_{kl} - 
g^{kl} \sqrt{g}
\mathring{\Gamma}\indices{^{i}_{kl}}
\, .
\end{align}
The result in a curved space with torsion is
\begin{equation}
\partial_t \rho (q,t)
= 
-\nabla_j  \left(
\left\{
u^j_+ (q,t)  + \nu K^{j}
\right\} \rho(q,t) \right) 
+ \nu \Delta \rho (q,t) 
\, . 
\label{eqn:fpeq_torsion}
\end{equation}
Here, $\nabla_i$ represents the covariant derivative and 
$\Delta = g^{ik} \nabla_i \partial_k$ 
is the connection Laplacian which reduces to the Laplace-Beltrami operator in the absence of torsion.
The contracted contortion tensor is defined by 
\begin{equation}
K_i \coloneq {K^k}_{i k } = 2 {S^k}_{i k} \eqqcolon 2 S_i \, ,
\end{equation}
which is induced by torsion and hence vanishes in a curved space without torsion.
The forward SDE and the corresponding FP equation show that spatial geometry, via the vielbein, modifies the noise term, thereby influencing the statistical behavior of particles. 
As discussed later, stochasticity caused by this noise forms the basis for interpreting quantum fluctuations. 
Notably, torsion affects the motion of fluctuating particles through its influence on the noise, even in the absence of spin, emphasizing its critical role in the quantum dynamics of spinless systems.

A stochastic trajectory can be continuous yet nowhere differentiable, even in the limit $\ud t \rightarrow 0$. 
That is, the left- and right-hand limits generally differ:
$\lim_{\ud t \rightarrow 0+} \ud \widetilde{q}^{i}(t)/\ud t \neq \lim_{\ud t \rightarrow 0-} \ud \widetilde{q}^{i}(t)/\ud t 
$.
To take the difference of the two limits into account, 
Nelson introduced two time derivatives \cite{NELSON}: 
one is the mean forward derivative, 
\begin{equation}
\uD_+ f(\widetilde{q}(t)) \coloneq \lim_{\ud t\rightarrow 0+} \uE \left[ \frac{f(\widetilde{q}(t+\ud t))-f(\widetilde{q}(t))}{\ud t} \Big| {\cal P}_t \right] \, ,
\label{eqn:mfd}
\end{equation}
and the other the mean backward derivative 
\begin{equation}
\uD_- f(\widetilde{q}(t)) \coloneqq \lim_{\ud t\rightarrow 0-} \uE \left[ \frac{f(\widetilde{q}(t+\ud t))-f(\widetilde{q}(t))}{\ud t} \Big| {\cal F}_t \right] \, .
\label{eqn:mbd}
\end{equation}
These expectation values are obtained from the conditional averages where ${\cal P}_t$ (${\cal F}_t$) indicates a set of trajectories fixing $\widetilde{q}(t^\prime)$ for $t^\prime \le t$ ($t^\prime \ge t$). 
The two mean derivatives play complementary roles in stochastic integration by parts. 
Specifically, for arbitrary stochastic processes $\widetilde{A}(t)$ and $\widetilde{B}(t)$, we find 
\begin{align}
& \int^b_a \ud s \, \uE\left[ \widetilde{B}(s) \uD_+ \widetilde{A}(s) \right] 
=
- \int^b_a \ud s \, \uE\left[ \widetilde{A}(s) \uD_- \widetilde{B}
(s) \right] \nonumber \\
& + \int^b_a \ud s\, \frac{\ud}{\ud s} \uE \left[ \widetilde{A}(s) \widetilde{B} (s)\right] \,.
\end{align}

Applying the above definition 
to the forward SDE (\ref{eqn:fSDE}), we can easily calculate the mean forward derivative,
\begin{equation}
\uD_+\widetilde{q}^{i}(t) = 
u^i_+ (\widetilde{q}(t),t) - \nu 
{\Gamma^{i}}_{k j} \, g^{k j} 
\, . \label{eqn:q_forwardderi}
\end{equation}
Here, ${\Gamma^{i}}_{k j}$ and $g^{k j} $ are evaluated 
at $\widetilde{q}(t)$ and 
thus are stochastic quantities themselves.
We cannot however calculate $\uD_- \widetilde{q}(t)$ from the forward SDE (\ref{eqn:fSDE}) since $\ud \widetilde{B}^{\hat{a}}_+(t)$ is defined only for $\ud t >0$.
We need to introduce an equivalent equation that evolves backward in time.

Introducing another vector field $u_-(q,t)$, this backward SDE is defined by
\begin{equation}
\ud \widetilde{q}^i (t) = 
u^i_- (\widetilde{q}(t),t) \ud t + \sqrt{2\nu} \tensor{e}{_{\hat{a}}^i} \circ \ud \widetilde{B}^{\hat{a}}_- (t) \, , \label{eqn:bSDE}
\end{equation}
where $\ud t < 0$, and $\ud \widetilde{B}^{\hat{a}}_- (t)$ denotes the inclination of an additional Wiener process, characterized by 
\begin{align}
\uE[\ud \widetilde{B}^{\hat{a}}_-(t)] &= 0 \, ,\\
\uE[\ud\widetilde{B}^{\hat{a}}_- (t) \ud\widetilde{B}^{\hat{b}}_- (t^\prime)] &= |\ud t| 
\delta^{\hat{a}\hat{b}}
\delta_{t t^\prime}\, .
\end{align}
The absolute value appearing in the second correlation is due to $\ud t<0$.
Using the backward SDE, we re-derive the FP equation. Since the backward SDE describes the time-reversed process of the forward SDE, the FP equations must be equivalent, which leads to the following consistency condition 
\cite{Matos20}:
\begin{equation}
u^i_+ (q,t) - u^i_-(q,t) = -2 \nu K^i + 2 \nu g^{ik} \partial_k \ln \rho (q,t) \, , \label{eqn:CC}
\end{equation}
leading to 
\begin{align}
\uD_- \widetilde{q}^{i}(t) 
&= 
u^i_- (\widetilde{q}(t),t) 
+ \nu {\Gamma^{i}}_{k j} \, g^{k j} \nonumber \\
&= 
u^i_+ (\widetilde{q}(t),t) + 2 \nu K^i \nonumber \\
&- 2 \nu g^{ik} \partial_k \ln \rho (\widetilde{q}(t),t) 
+ \nu {\Gamma^{i}}_{k j} \, g^{k j} 
\, .\label{eqn:q_backwardderi}
\end{align}

Using the consistency condition, 
we find that the two FP equations reduce to the same equation of continuity, 
\begin{align}
\partial_t \rho (q,t) 
=- \nabla_j (\rho (q,t) v^{j} (q,t) ) \, , \label{eqn:eq_conti}
\end{align}
where 
\begin{equation}
v^i (q,t):= \frac{u^i_+ (q,t)+ u^i_- (q,t)}{2} \, . \label{eqn:del_v}
\end{equation}
We call $v^i(q,t)$ the particle velocity field.
In contrast to the two vector fields $u^i_\pm(q,t)$, 
$v^i(q,t)$ is defined as the velocity of the probability current. 
We emphasize that this deterministic field must be distinguished from the velocity of an individual particle on its trajectory  
$\widetilde{q}(t)$. 
Indeed, a unique velocity for the particle cannot be defined, a direct consequence of the non-differentiable nature of its trajectory. 
This non-differentiability is mathematically manifested by the inequality of the mean forward and backward velocities (given by Eqs. (\ref{eqn:q_forwardderi}) and (\ref{eqn:q_backwardderi})). 
This inherent ambiguity in defining an instantaneous particle velocity is precisely the origin of the uncertainty relation within the SVM framework \cite{KOIDE20181472,GAZEAU20,Matos20}.

\section{GENERALIZED QUANTUM DYNAMICS WITH TORSION}
\label{sec:nonlinearSCH}

\subsection{Setup of the Lagrangian}

We now develop a generalized scheme of SVM to derive the quantum dynamics of a spinless particle in a curved space with torsion. 
We start from the classical Lagrangian for a non-relativistic particle of mass $m$
moving in a curved space with torsion,
\begin{equation}
L = \frac{m}{2} g_{ij} \dot{q}^{i} \dot{q}^j - V (q) \, ,
\end{equation}
where $V(q)$ represents the potential energy and
$\dot{q}^{i} := \ud q^{i}/\ud t$.
Note that this Lagrangian does not explicitly depend on torsion. 
The variation of this Lagrangian with respect to the generalized coordinates $q^i$ gives the well-known equation of motion:
\begin{equation}
\ddot{q}^i + \mathring{\Gamma^{i}}_{j k} \dot{q}^j \dot{q}^k 
= - \frac{1}{m} g^{ik} \partial_k V (q) \label{eqn:cla_eom_curve1}\, ,
\end{equation}
where $\ddot{q}^i = \ud^2 q^{i}/ \ud t^2$ and $\mathring{\Gamma^{i}}_{j k}$ is the Levi-Civita connection. For a free particle ($V=0$), this is the geodesic equation.

Alternatively, the same Lagrangian can be expressed in local Cartesian coordinates,
\begin{equation}
L = \frac{m}{2} \eta_{\hat{a}\hat{b}} \dot{q}^{\hat{a}} \dot{q}^{\hat{b}} - V \left( q \right) \, , \label{eqn:curve_cla_lag2}
\end{equation}
where $\eta_{\hat{a}\hat{b}} = {\rm diag} ( 1,1,\cdots )$. The variation with respect to $q^{\hat{a}}$ gives the simpler equation:
\begin{equation}
\ddot{q}^{\hat{a}}= - \frac{1}{m}\eta^{\hat{a}\hat{b}} \partial_{\hat{b}} V  \, .
\end{equation}
The resulting equation of motion must be independent of the coordinate system used to express the Lagrangian. 
To check this consistency, we transform the local equation of motion back to generalized coordinates. 
Using the relation $\dot{q}^i = \tensor{e}{_{\hat{a}}^i} \dot{q}^{\hat{a}}$ and the horizontal lift of the vielbein,
$\tensor{\ud e}{_{\hat{a}}^i} 
= -  {\Gamma^{i}}_{j k}  \tensor{e}{_{\hat{a}}^j}   \ud q^k(t)$ 
which is the deterministic version of Eq.\ (\ref{eqn:de}), 
%is governed by the full affine connection $\Gamma$ as Eq.\ (\ref{eqn:de}), 
we obtain:
\begin{align}
\ddot{q}^i 
&= \left(\frac{\ud}{\ud t} \tensor{e}{_{\hat{a}}^i}\right) \dot{q}^{\hat{a}} + \tensor{e}{_{\hat{a}}^i} \ddot{q}^{\hat{a}} \nonumber \\
%&= - {\Gamma^{i}}_{j k}\dot{q}^k (\tensor{e}{_{\hat{a}}^j} \dot{q}^{\hat{a}}) - \frac{1}{m} g^{ik} \partial_k V \nonumber \\
&= - {\Gamma^{i}}_{j k}\dot{q}^k \dot{q}^{j} - \frac{1}{m} g^{ik} \partial_k V 
\label{eqn:cla_eom_curve2}\, .
\end{align}
At first glance, this result (\ref{eqn:cla_eom_curve2}) appears to contradict Eq.\ (\ref{eqn:cla_eom_curve1}), as it involves the full affine connection $\Gamma$ instead of the Levi-Civita connection $\mathring{\Gamma}$. 
This is a well-known fundamental issue, often referred to as the problem of disagreement between autoparallels and geodesics in theories with torsion \cite{Hehl1}.
While this inconsistency might be resolved by considering a non-minimal coupling of torsion in the Lagrangian, there is no universally established procedure for this, especially for spinless particles. 
Therefore, instead of modifying the Lagrangian, we impose a restriction on the property of torsion itself. 
The consistency between Eqs.\ (\ref{eqn:cla_eom_curve2}) and (\ref{eqn:cla_eom_curve1}) 
is achieved if the contortion tensor term $K\indices{^{i}_{jk}}\dot{q}^j \dot{q}^k$ vanishes. 
Since $\dot{q}^j \dot{q}^k$ is symmetric in $j,k$, 
this is satisfied if the torsion tensor $S\indices{^{i}_{jk}}$ is {\it totally antisymmetric}.
In this situation, the torsional contribution to the equation of motion disappears, 
and the two equations become identical. Consequently, in the remainder of this paper, 
we will exclusively focus on curved spaces with totally antisymmetric torsion.

This assumption of the totally antisymmetry is also plausible from 
a cosmological perspective. 
It is known that employing the cosmological principle together with
energy-momentum conservation 
and standard cosmological expansion restricts
the form of torsion to the case of a totally antisymmetric tensor \cite{vandeVenn22,Tsamparlis79}.
The totally antisymmetrc torsion is furthermore 
preferred from a metric-affine gravity standpoint as it allows for a
continuous limit of vanishing torsion, thereby recovering GR \cite{Fabbri14}.

To implement the stochastic variation, 
$\dot{q}^{\hat{a}}$ should be replaced with 
the corresponding stochastic quantity.
In the context of quantization, it is known that the kinetic term is replaced with its symmetric average over the mean forward and backward derivatives \cite{Matos20,Koide11,KOIDE20192713}: 
\begin{equation}
\eta_{\hat{a}\hat{b}} \dot{q}^{\hat{a}} \dot{q}^{\hat{b}} \longrightarrow 
\eta_{\hat{a}\hat{b}} \frac{(\uD_+ \widetilde{q}^{\hat{a}})(\uD_+ \widetilde{q}^{\hat{b}}) + (\uD_- \widetilde{q}^{\hat{a}})(\uD_- \widetilde{q}^{\hat{b}})}{2} \, .
\end{equation}
This symmetric replacement ensures time-reversal invariance in the stochastic Lagrangian. Violation of this symmetry leads to a viscous effect \cite{Koide11,Matos20,Koide_2022_JSTAT,Koide_2020}.
By employing a Lagrangian that violates time-reversal symmetry, SVM can provide a variational foundation for more general classical hydrodynamical equations, such as the Navier-Stokes-Fourier equation, the bivelocity hydrodynamics proposed by Brenner, and the Navier-Stokes-Korteweg equation \cite{Matos20}.

In this paper, we require that the standard Schr\"{o}dinger equation is reproduced in the absence of torsion  and thus this symmetric requirement is the most natural starting point for our framework.
Therefore, the corresponding stochastic Lagrangian is given by 
\begin{align}
&L_{sto} (\widetilde{q}, \uD_+ \widetilde{q} , \uD_- \widetilde{q}) \nonumber \\
&= \frac{m}{4} \eta_{\hat{a}\hat{b}} \{ (\uD_+ \widetilde{q}^{\hat{a}})(\uD_+ \widetilde{q}^{\hat{b}}) + (\uD_- \widetilde{q}^{\hat{a}})(\uD_- \widetilde{q}^{\hat{b}}) \} - V (\widetilde{q})\, .
\end{align}

\subsection{Non-linear Schr\"{o}dinger equation}

An individual particle trajectory $\widetilde{q}(t)$ follows Brownian motion and is thus not fully controllable. 
As is evident from Eq.\ (\ref{eqn:fSDE}), 
its trajectory cannot be determined solely by the choice of the vector field $u^i_+ (q,t)$. 
Consequently, the variational principle can only be meaningfully applied to control the average behavior of an ensemble of trajectories. 
For this reason, we apply the variation to the ensemble-averaged action,
\begin{equation}
I_{sto} \left[ \widetilde{q} \right]= \int_{t_i}^{t_f} \ud t \,\uE \left[ 
L_{sto} (\widetilde{q}, \uD_+ \widetilde{q} , \uD_- \widetilde{q}) \right] \, , \label{eqn:sto_action}
\end{equation}
defined over an arbitrary time interval $t_i \le t \le t_f$. 
The variation of the particle trajectory is defined by 
\begin{equation}
\widetilde{q}^{\hat{a}} (t) \longrightarrow \widetilde{q}^{\hat{a}} (t) + \delta f^{\hat{a}} (\widetilde{q} (t),  t) \, , 
\end{equation}
where the infinitesimal, smooth function satisfies 
the following boundary conditions:
\begin{equation}
\delta f^{\hat{a}} (q,  t_i) = \delta f^{\hat{a}} (q,  t_f) = 0 \, .
\end{equation}

For example, the kinetic term is calculated as 
\begin{align}
&\delta \int_{t_i}^{t_f} \ud t \uE \left[ 
\eta_{\hat{a}\hat{b}}  (D_\pm q^{\hat{a}})(D_\pm q^{\hat{b}})
\right] \nonumber \\
&= 
2 \int_{t_i}^{t_f} dt \uE \left[ \eta_{\hat{a}\hat{b}}  
u_\pm^{\hat{a}} (D_\pm \delta f^{\hat{b}} )
\right] \nonumber \\
&= 
- 2 \int_{t_i}^{t_f} \ud t  \uE \left[ 
 \delta f^{\hat{b}} 
g_{ij}   e^j_{\hat{b}} e^i_{\hat{a}}
D_\mp u_\pm^{\hat{a}}   
\right] \nonumber \\
&= 
- 2 \int_{t_i}^{t_f} \ud t   \uE \left[ 
g_{ik} \delta f^{k}    
\{
\partial_t u^i_\pm + u^j_\mp \nabla_j u^i_\pm \mp \nu \Delta u^i_\pm 
\}
\right] \, .
\end{align}

Requiring the stochastic variation of Eq.\ (\ref{eqn:sto_action}) to vanish for any choice of $\delta f(q,t)$ 
and 
stochastic events $\ud \widetilde{B}^{\hat{a}}_\pm (t)$, 
%$\delta I_{sto} [\widetilde{q}]=0$, 
we find that the particle velocity field $v^{i}(q,t)$ is determined by the following hydrodynamical equation,  
\begin{align}
& \partial_t v^i + v^k \nabla_k v^i 
+ \frac{1}{m}
g^{ik}\partial_k V \nonumber \\
& =
 \nu^2  g^{ik}
\left\{
2\nabla_k \left(\frac{1}{\sqrt{\rho}}\Delta\sqrt{\rho} \right)
 -  R_{j k} g^{j l}\partial_l \ln \rho \right\}
+ \nu^2 \Sigma^i   \, ,\label{eqn:val_eq_v}
\end{align}
where ${R_i}^k = g^{kj} R_{ij} = g^{jk} {R^l}_{jkl}$ is the Ricci tensor and 
\begin{align}
\Sigma^i
&= 
 2 \ g^{ij}g^{lm} \left\{
-(\partial_m \ln \rho) {S^n}_{jl} (\partial_n \ln \rho) 
+ {S^n}_{lj}  \nabla_n \partial_m \ln \rho  \right. \nonumber \\
& \left. - \nabla_m ({S^n}_{jl} \partial_n \ln \rho) 
\right\}  -  
K^j \nabla_j K^i \nonumber \\
& + \Delta K^i
+ g^{jm} (\partial_m \ln \rho) \nabla_j K^i 
+ K^j \nabla_j g^{ln}\partial_n \ln \rho \, . \label{eqn:sigma}
\end{align}
In this derivation, we have used a general relation satisfied by an arbitrary vector, 
\begin{align}
[ \nabla_j,  \nabla_k] A^i 
&= {R^i}_{l j k} A^l 
- 2 {S^l}_{k j}  \nabla_l A^i \, ,
\end{align}
and the following relations,
\begin{widetext}
    \begin{align}
        g^{kl} (\partial_l \ln \rho) \nabla_k g^{ij}(\partial_j \ln \rho)
        &= \frac{1}{2} g^{ij} \nabla_j (g^{kl} (\partial_l \ln \rho) (\partial_k \ln \rho))
        - 2 g^{kl}g^{ij} (\partial_l \ln \rho) {S^\rho}_{jk} \partial_\rho \ln \rho \, ,\\
        g^{km}g^{ij}\nabla_k \nabla_m \partial_j \ln \rho 
        &= g^{ij}\nabla_j \Delta \ln \rho -g^{ij} R_{k j} g^{kl}\partial_l \ln \rho
        + 2 g^{ij}  g^{kl} ( {S^m}_{kj}  \nabla_m \partial_l \ln \rho  -\nabla_l ({S^m}_{jk} \partial_m \ln \rho) ) \, ,\\ 
        \frac{1}{\sqrt{\rho}} g^{ij} \nabla_i \partial_j \sqrt{\rho}
        &=
        \frac{1}{2} g^{ij} 
        \left(
        \nabla_i \partial_j \ln \rho + \frac{1}{2}(\partial_j \ln \rho) (\partial_i \ln \rho)
        \right) \, .
    \end{align}
\end{widetext}

In the limit of vanishing noise, $\nu \to 0$, Eq.\ (\ref{eqn:val_eq_v}) reduces to Newton’s second law. 
The geodesic equation then follows from $\ud q^{i}(t)/\ud t = v^{i} (q(t),t)$ and $\ud v^{i} (q(t),t)/\ud t = (\partial_t + v^k(q(t),t) \partial_k)v^{i} (q(t),t)$.
As discussed in Refs.\ \cite{YASUE81,ZAMBRINI,Koide11,KOIDE20192713,Matos20} and shown later in Eq.\ (\ref{eqn:log-schriedinger}), the noise intensity $\nu$ is determined by the Planck constant, reproducing quantum mechanics.
Consequently, all terms on the right-hand side in Eq.\ (\ref{eqn:val_eq_v}) stem from quantum effects.
The first term on the right-hand side corresponds to the gradient of Bohm’s quantum potential, 
while the second reflects quantum fluctuations and spatial curvature, as shown in Ref.\  \cite{KOIDE20192713}. 
The newly introduced term $\Sigma^{i}$ captures the interaction between quantum fluctuations and torsion.
The extended SVM formulation and the hydrodynamic equation (\ref{eqn:val_eq_v}), developed here, 
are key contributions of this study.

While the properties of the torsion tensor have not been explicitly used thus far, we can simplify Eq.\ (\ref{eqn:val_eq_v}) by using its total antisymmetry. For $D=3$, this is expressed as:
\begin{equation}
S_{abc} = s(t)\epsilon_{abc} \, ,   
\end{equation}
where $s(t)$ is a scalar field and $\epsilon_{abc}$ the $3-$dimensional Levi-Civita tensor. 
This follows directly from the Hodge duality relation
$\tensor{\star S}{_{a_1 \ldots a_{D-3}}} = \tensor{\epsilon}{^{b_1 b_2 b_3}_{a_1 \ldots a_{D-3}}}\tensor{S}{_{b_1 b_2 b_3}}/3!$ for $D=3$.
Here, the Hodge star is given by the scalar field $s(t)$.
In this formulation, torsion
satisfies $S_i = 0$ and becomes equivalent to the contortion
tensor, i.e., $K_{ijk} = S_{ijk}$.

Our velocity equation (\ref{eqn:val_eq_v}) is then simplified as
\begin{align}
&    \partial_{t}v^i + v^k \mathring{\nabla}_k v^i 
   + \frac{1}{m}g^{ik}\partial_k V 
   -2 \nu^2 g^{ik}
    \partial_k \left(\frac{1}{\sqrt{\rho}}\mathring{\Delta}\sqrt{\rho}\right)  \nonumber \\
&= 
    - \nu^2 g^{ik}\biggl\{ 2 (\partial_k \ln\rho)s^2 + \mathring{R}_{l k}
    g^{l j}\partial_{j}\ln\rho 
    + (\partial_j \ln\rho) \mathring{\nabla}_l {S^{l j}}_{k} 
    \biggr\} \, , \label{eqn:v_tastorsion}
\end{align}
where $\mathring{R}_{i k}$, $\mathring{\nabla}_i$ and $\mathring{\Delta}$ are the Ricci tensor, the covariant derivative and the Laplace-Beltrami operator, based soly on the torsion-free Levi-Civita connection, respectively.
Thus all torsion effects are induced by the right-hand side of Eq.\ \eqref{eqn:v_tastorsion}. 
In this derivation, we used 
\begin{align}
R_{ik} &= \mathring{R}_{ik} - \mathring{\nabla}_j {S^{j}}_{ik}
    + {S^{j}}_{l k}{S^{l}}_{i j} \, , \\    
{S^{b}}_{a k}{S^{a i}}_{b} &= 2 s^2\delta^i_k \, .
\end{align}

To recast the hydrodynamical equation into the form
of the Schr\"{o}dinger equation and thereby transitioning to a quantum-mechanical description of our system,
we further consider a flat space ($\mathring{R}_{ik}=0$) with torsion.
A complex function $\Psi (q,t)$ is introduced:
\begin{equation}
\Psi (q,t) = \sqrt{\rho(q,t)} e^{\ii \theta (q,t)} \, , \label{eqn:wavefunction}
\end{equation} 
where $\theta (q,t)$ is defined by
\begin{equation}
v^{i} (q,t) = 2 \nu g^{ij} \partial_j \theta(q,t) \, . \label{eqn:phase}
\end{equation}
Choosing $\nu = \hbar/(2m)$, the equation for the complex function, the wave function, 
is given by the nonlinear Schr\"{o}dinger equation,
\begin{equation}
\ii \hbar \partial_t \Psi = 
\left[
- \frac{\hbar^2}{2m} \Delta
+ V + \frac{\hbar^2}{2m} s^2 \ln |\Psi|^2
\right] \Psi \, , \label{eqn:log-schriedinger}
\end{equation}
reducing to the Sch\"{o}dinger equation in the limit of vanishing torsion.
This equation validates our conjecture that torsion and quantum fluctuations fundamentally modify spinless quantum mechanics, by introducing nonlinearity.
This represents another significant result of this work.

\subsection{Nonlinearity and constraints for the magnitude of torsion}

Nonlinear modifications to the Schr\"{o}dinger equation have been extensively studied in various contexts, 
including soliton dynamics, wave function collapse models, and alternative formulations of quantum mechanics \cite{kostin,Bialynicki75,Bialynicki76,kibble,dekker,DIOSI1987377,PhysRevA.40.1165,PhysRevLett.62.485,WEINBERG1989336,PhysRevA.40.3387,Polchin,penrorse,CZACHOR19971,PhysRevA.57.4122,RevModPhys.85.471,CZACHOR2002139,KOIDE20181472,dieter,al}.
Our findings extend these phenomenological approaches naturally.
For instance, logarithmic nonlinear terms emerge in the phenomenological quantization of classical dissipative systems, although they often feature purely imaginary coefficients \cite{kostin,dekker,PhysRevA.40.3387,KOIDE20181472,dieter}.

The nonlinear extension of quantum mechanics in a flat space has been studied  \cite{Bialynicki75,Bialynicki76}, where the log-nonlinear term is identified as the most viable candidate. 
Our equation (\ref{eqn:log-schriedinger}) matches this result, preserving essential quantal properties such as the Born rule and Ehrenfest theorem, thus ensuring classical-quantum correspondence. 
In contrast to Refs.\ \cite{Bialynicki75,Bialynicki76}, which treat the logarithmic coefficient as a free parameter, our formulation derives it explicitly from torsion. 
However, the positive sign of our nonlinear term, as suggested by Refs.\ \cite{Bialynicki75,Bialynicki76}, indicates that the resulting nonlinear equation might lack stationary solutions, raising open questions about the stability of the wave function.

Furthermore, it gives rise to other well-studied theoretical implications, including: (i) asymptotic classicality, where initially non-orthogonal states can evolve to become perfectly distinguishable \cite{mielnik}; (ii) the potential for superluminal signaling \cite{gisin}; and (iii) the violation of the mixture equivalence principle, wherein distinct physical preparations of the same density matrix evolve differently \cite{pttr-6kj7}. Crucially, since the strength of our nonlinear term is governed by the torsion magnitude, the experimental non-observation of these phenomena provides a powerful method to constrain its value, implying that the magnitude of $s(t)$ must be extremely small.

For instance, in a flat space, torsion could lead to small shifts in the hydrogen atom's energy levels, although these shifts are likely negligible compared to the Coulomb potential. 
By evaluating potential energies at the Bohr radius $R_B$ 
with $\ln \left |\Psi (R_B) \right|^2 \sim 1$, we derive an upper bound for the torsion magnitude:
\begin{equation}
s \ll \sqrt{2\alpha \frac{m_e}{\hbar^2 R_{B}}} \approx 5.3 \times 10^{-3} \text{ MeV} \sim 10^{-2} \,m_e ,
\end{equation}
where $\alpha$ is the fine-structure constant and $m_e$ is the electron mass.
Our derived bound is not in contradiction with the much more stringent experimental constraints from Ref.\ \cite{PhysRevLett.100.111102}, 
which suggest $s \leq 10^{-31}~\text{GeV}$ under the assumption of Lorentz symmetry violation.

\section{Concluding remarks}
\label{sec:conclusion}

In this study, we investigated the previously uncharted quantum dynamics of non-relativistic spinless particles in the presence of torsion by introducing an extended stochastic variational method (SVM). 
Our central finding is that torsion's influence extends beyond spin, giving rise to a logarithmic nonlinearity in the Schr\"{o}dinger equation. 
This effect, a direct consequence of our stochastic variational approach, originates from quantum fluctuations modulated by torsion, proposing a new physical mechanism for nonlinear quantum phenomena.

In our view, quantization is an extended variational principle wherein quantum fluctuations, manifested as Brownian motion, are inevitably shaped by the underlying geometric structure. The nonlinear term derived in this paper is a direct physical consequence of these quantum fluctuations being distorted by torsion. Therefore, within our framework, this nonlinearity is not a phenomenological addition but an unavoidable consequence. It should thus be considered a fundamental modification of physical law, distinct from the phenomenological nonlinearities postulated in collapse models to explain wave function collapse.

Motivated by similar considerations, Rapoport also investigated the effect of torsion on the Schr\"{o}dinger-like equation \cite{rapoport}. 
There are two fundamental differences between Rapoport's approach and ours.
The first difference lies in the treatment of torsion.
Rapoport introduces a correspondence where the trace-torsion (the vector component of torsion) maps to the drift term of Brownian motion. 
Our work, by contrast, employs a totally anti-symmetric torsion, for which the trace is necessarily zero.
The second difference concerns the foundational premise. 
Whereas we derive quantum mechanics employing a variational principle, Rapoport’s approach is predicated on a reciprocal process, postulating that its transition probability is described by a diffusion equation modified by torsion. 
This reciprocal process, originally introduced by Schr\"{o}dinger from a classical probabilistic perspective \cite{reciprocal,rec_koide,rec_review}, is a fundamentally real-valued framework. 
As a result, Rapoport must artificially construct the complex wave function a posteriori from two real functions.
Therefore, while Rapoport suggested that torsion could be a source of nonlinearity, the work does not demonstrate that any specific form of it is a necessary consequence of the theory.

Equation (\ref{eqn:val_eq_v}), if we ignore torsion and spatial curvature, coincides with the Madelung equation, which is known as the hydrodynamical representation of the Schr\"{o}dinger equation. 
Since Bohmian mechanics is a rewriting of the Madelung equation for fluid elements, the results of SVM and Bohmian mechanics are related. However, the results of the two approaches are generally not identical. 
A key difference is that Bohmian mechanics identifies the particle's trajectory with that of the fluid element, meaning two trajectories starting from different initial positions will never cross. 
In contrast, the Brownian trajectories used in SVM can and do intersect. 
Furthermore, in Bohmian mechanics, the uncertainty principle is explained by incorporating the contribution of the quantum potential into the standard deviation of momentum. In SVM, on the other hand, the uncertainty principle is interpreted as arising from the ambiguity in the definition of momentum 
due to the non-differentiability of the Brownian particle's trajectory \cite{KOIDE20181472,GAZEAU20,Matos20}.

In the SVM quantization, 
the form and the coefficient of the nonlinear term 
are determined by a chosen Lagrangian.
A time-dependent Lagrangian that violates time-reversal symmetry gives rise to a different nonlinear term with a purely imaginary coefficient \cite{KOIDE20181472,PhysRevA.40.3387}.
Meanwhile, choosing a hyperspherical surface yields a log-nonlinear term with a negative real coefficient associated with the Ricci scalar \cite{KOIDE20192713}.
Notably, 
the Gross-Pitaevskii equation is obtained by the SVM quantization of an ideal fluid Lagrangian \cite{Koide11}.
These examples underscore the adaptability 
of SVM as a natural framework for extending quantum theory nonlinearly.

Although similar torsion-induced nonlinearities are observed in the Dirac equation \cite{Hehl1,zecca,POPLAWSKI201073,POPLAWSKI2013575,Fabbri}, the mechanism here uniquely affects the Schr\"{o}dinger equation rather than emerging via gravitational potentials as in the Schrödinger-Newton equation \cite{PhysRevD.98.104027}. Future investigations will address torsion’s impact on the uncertainty principle \cite{KOIDE20181472,GAZEAU20, Koide_2022_JSTAT}, black hole thermodynamics \cite{ONG2018217}, and further extend the SVM method to include relativistic fields \cite{Koide13}. Investigating these questions promises to deepen our understanding of torsion’s significance in quantum mechanics and gravitational physics.

\begin{acknowledgments}
T.K. thanks for the fruitful discussion with T. Kodama and the theory groups
of the Institute for Theoretical Physics in the Johann Wolfgang Goethe University
and the Frankfurt Institute for Advanced Studies (FIAS). 
T.K. acknowledges the financial support by CNPq (No.\ 305654/2021-7;304504/2024-6). T.K. and A.vdV. are grateful for support from the Fueck-Stiftung.
A part of this work has been done under the project INCT-Nuclear Physics and Applications (No.\ 464898/2014-5).
\end{acknowledgments}

\bibliography{torsionSVM}

\end{document}